\documentclass[12pt]{article}

\pdfoutput=1 

\newcommand{\blind}{0}

\usepackage{amsmath,amsfonts,amssymb}
\usepackage{latexsym,textcomp,cancel}
\usepackage{graphicx}
\usepackage{natbib}
\usepackage{appendix}
\usepackage{url,color}
\usepackage{multirow, rotating}
\usepackage{cancel}
\usepackage{enumerate}

\newcommand{\bfalpha} {\boldsymbol{\alpha}}

\newcommand{\bftheta} {\boldsymbol{\theta}}

\newcommand{\bfTheta} {\boldsymbol{\Theta}}

\newcommand{\bfx} {\mathbf{x}}

\newcommand{\bfX} {\mathbf{X}}
\newcommand{\bfY} {\mathbf{Y}}

\renewcommand{\Pr}{\mathsf{Pr}}

\DeclareMathOperator{\logit}{logit}

\DeclareMathOperator{\argmax}{argmax}
\DeclareMathOperator{\sign}{sign}

\begin{document}

\def\spacingset#1{\renewcommand{\baselinestretch}%
{#1}\small\normalsize} \spacingset{1}

\if0\blind
{
  \title{\bf Investigating Competition in Financial Markets: A Sparse Autologistic Model for Dynamic Network Data}
  \author{Brenda Betancourt
  \hspace{.2cm}\\
    Department of Statistics, UC Santa Cruz\\
     and \\
    Abel Rodr{\'i}guez  \\
    Department of Statistics, UC Santa Cruz\\
    and \\
    Naomi Boyd  \\
    Department of Finance, West Virginia University}
  \maketitle
} \fi

\if1\blind
{
  \bigskip
  \bigskip
  \bigskip
  \begin{center}
    {\LARGE\bf Bayesian Fused Lasso regression for dynamic binary networks}
\end{center}
  \medskip
} \fi

\bigskip

%

%


\begin{abstract}
We develop a sparse autologistic model for investigating the impact of diversification and disintermediation strategies in the evolution of financial trading networks. In order to induce sparsity in the model estimates and address substantive questions about the underlying processes the model includes an $L^1$ regularization penalty. This makes implementation feasible for complex dynamic networks in which the number of parameters is considerably greater than the number of observations over time.  We use the model to characterize trader behavior in the NYMEX natural gas futures market, where we find that disintermediation and not diversification or momentum tend to drive market microstructure.
\end{abstract}

\noindent%
\vfill

\newpage
\spacingset{1} 

\section{Introduction}\label{sec:intro}

Historically, the empirical study of financial markets has emphasized the behavior of aggregate measures such as price discovery or transaction volumes.  However, the challenges that come from the rise of electronic and automated trading have highlighted the need to study patterns of individual market transactions (the so-called market microstructure) in order to understand the mechanisms that underpin price formation.  

Financial Trading Networks (FTNs), which are directed graphs in which nodes correspond to traders operating in a financial market and edges/weights represent pairwise buy-sell transactions among them that occur within a period of time, are becoming a popular tool for studying the complexity associated with modern financial markets. Indeed, FTNs contain key information about patterns of order execution in order-driven markets, which can in turn provide important insights into the functioning of the market.  Empirical work on trading networks so far has focused on studying the evolution of summary statistics such as degree distributions, average betweenness and clustering coefficients  and their relationship with market variables such as the volatility of returns (e.g., see \citealp{AdBrHaKi10}).  Model-based approaches are quite rare, one exception is \cite{BetaRodBoyd15}, where a hidden Markov model for dynamic network data is introduced to identify change points in the underlying market microstructure. 

In this paper we are interested in using FTNs to investigate whether traders in financial markets engage in strategic behaviors such as diversification and disintermediation as part of their long term trading strategies. To accomplish this goal we extend the notion of an autologistic model to directed binary network data. In the much simpler case of a binary time series $y_1, y_2, \ldots, y_T$ with $y_t \in \{ 0, 1 \}$, a first order autologistic model with parameters $\alpha$ and $\xi$ assumes that    
%
%
\begin{align}\label{eq:simple_autologistic}
\logit \Pr(y_t = 1 \mid y_{t-1}, \ldots, y_1) &= \eta_t
  =  \alpha + \xi y_{t-1}
\end{align}
where $\logit x = \log \{ x/(1 -x) \}$ and the unknown parameters $\alpha$ and $\xi$ control the structure of the temporal dependence (in particular, note that $\xi = 0$ implies that the observations are independent and identically distributed).  This implies that
\begin{align*}
p(y_2, \ldots, y_T \mid y_1, \alpha, \xi) =  
\prod_{t=2}^{T} 
\frac{ \exp\left\{ y_t (\alpha + \xi y_{t-1}) \right\} }
{ 1 + \exp\left\{ \alpha + \xi y_{t-1} \right\}} .
\end{align*}
Autologistic models for spatio-temporal binary data have been discussed in \cite{ZhuZheCar08} and \cite{ZhengZhu08}.  However, these models for spatio-temporal data cannot be directly applied to the network time series data discussed in this paper because they are not designed to account for common features of directed network data such as reciprocity (e.g., the tendency of nodes in the network to consistently respond to a positive action with another positive action) and transitivity (e.g., the tendency of nodes to interact if they share links with a common third party).  In contrast, the class of models we introduce in this paper are specifically designed to account for these features, and its parameters have a direct interpretation in terms of network properties.


The autologistic models developed in this paper are special cases of the so-called $p_1$ models of \cite{Holland}, which have been extended to dynamic settings in \cite{BaCa96}, \cite{GoldZhAi09}, and \cite{Kolac09}, among others.  However, our approach differs from these by directly incorporating the observed network links at previous time points rather than relying on summary network statistics. Our approach is also loosely related to the temporal version of the Exponential Random Graph model (ERGM) introduced in \citet{HanFuXi10} and further developed in \citet{CranDes11} and \citet{SnijSteBunt10}.  However, our approach allows for networks effects to be different for each pair of nodes, leading to additional expressive power and richer interpretation.  Other relevant approaches for modeling network data include the dynamic version of the latent space model of \cite{Hoff2} developed by \citet{SarkarMoore05} and \citet{SewChen15}, the work of \citet{XiFuSo10} presenting the temporal extension of the stochastic blockmodel first introduced in \citet{Airoldi}, the work of \citet{HuangLin09}, who present an autoregressive integrated moving average model and combine it with link occurrence scores based on similarity indices of network topology measures (e.g. Adamic-Adar coefficient, Katz index), and \citet{BlissFrank14}, who propose a method based on similarity indices and node attributes (e.g. common neighbors and preferential attachment) together with a covariance matrix adaptation evolution strategy for link prediction in networks with a large number of nodes.

One challenge for statistical inference for the class of models we discuss in this paper is that the number of parameters is large and grows linearly with the number of nodes in the network.  Indeed, unless the system is observed for a very long time, the number of parameters in our autologistic model will typically be much larger than the number of available observations. To deal with this challenge we employ $L^1$ regularization which is particularly useful to reduce the number of parameters in the $ p > N $ case. The use of an $L^1$ penalty leads to sparse solutions in which a large number of model coefficients are set to zero, allowing us to address substantive questions about the type of processes driving the evolution of the network. In addition, while performing short-term predictions is a secondary goal for our model, prediction accuracy can sometimes be improved by shrinking the model coefficients \citep{Tibshi96}. Computation of the procedure is carried out using a coordinate descent method on a surrogate quadratic approximation for an $L^1$ regularized multinomial likelihood (e.g., see \citealp{FriHaTib10}).

The methodology described in this paper is illustrated using a dataset on transactions in the NYMEX natural gas futures market that took place between January 2005 and December 2008. This dataset was previously analyzed in \cite{BetaRodBoyd15} using a hidden Markov model to identify points of structural change in the market.  In contrast, the analysis in this paper suggests that disintermediation effects tend to be the most important drivers of network evolution, a pattern that was previously unknown and is consistent with competitive markets. The remainder of the paper is organized as follows:  Section \ref{sec:model} describes our model and discusses some of its properties.  Section \ref{sec:computation} describes our computational algorithm and some of the properties of the estimators.  Section \ref{sec:applications} discusses two illustrations, one based on simulated data and a second one that focuses on real trading networks from the natural gas futures market on the New York Mercantile Exchange (NYMEX). Finally, Section \ref{sec:discussion} presents a short discussion.

\section{Data}\label{se:data}

The dataset we analyze in this paper consists of individual transaction records from the New York Mercantile Exchange (NYMEX), a commodity futures exchange owned by the Chicago Mercantile Exchange. Commodities traded on NYMEX include coal, electricity, palladium uranium, and natural gas, among others.  Our analysis focuses on proprietary trades (i.e., transactions carried out by traders for their own accounts rather than on their client's behalf) in the natural gas futures market covering
the period from January 2005 to December 2008. During this period, the NYMEX natural gas market operated as an open outcry market until September 5, 2006, and as a hybrid market that included electronic trading (conducted via the CME Globex platform) after that date.  Over 900 unique traders participated in the market during the period under study;  however, the vast majority of the traders participated in the market only sporadically.  Furthermore, we have no access to information about whether a specific trader entered or left the market at a given point in time.  Hence, we focus our analysis on 71 large traders identified as being present in the market (although not necessarily active) during the whole period.  From the original transaction data we construct a sequence of weekly binary FTNs by setting the entries of the adjacency matrices to $y_{i,j,t} =1$ if there was at least one transaction in which trader $i$ sold a contract to trader $j$ during week $t$.  

\section{Modeling approach}\label{sec:model}

Consider a sequence of $T$ binary directed networks, each one observed over a common set of $n$ nodes.  The adjacency matrix of the network at time $t$ is therefore an $n \times n$ binary matrix $\bfY_{t} = [ y_{i,j,t} ]$, where $y_{i,j,t} = 1$ if there is a link directed from node $i$ to node $j$ at time $t$, and $y_{i,j,t} = 0$ otherwise. We adopt the convention $y_{i,i,t} \equiv 0$ so that there are no loops within the network.  In the illustration we discuss in Section \ref{sec:applications}, the nodes in the network correspond to traders in NYMEX natural gas futures market, so that $y_{i,j,t} = 1$ if trader $i$ sold a contract to trader $j$ at least once during week $t$.

We consider an extension of \eqref{eq:simple_autologistic} in which the pairs $\{ (y_{i,j,t} , y_{j,i,t}) : i< j \}$ are assumed conditionally independent given the history of the network, and each pair $(y_{i,j,t} , y_{j,i,t})$ is modeled using a logistic model of the form
\begin{multline}\label{eq:model}
p\left(y_{i,j,t},y_{j,i,t} \mid \bfY_{t-1} \right) =   
 \exp  \big\{ \eta_{i,j,t,1} y_{i,j,t}
 + \eta_{i,j,t,2} y_{j,i,t} + \eta_{i,j,t,3} y_{i,j,t}y_{j,i,t} \\
 - C\left(\eta_{i,j,t,1}, \eta_{i,j,t,2},\eta_{i,j,t,3} \right) \big\}  ,
\end{multline}
where the normalizing factor is given by
\begin{multline*}
C\left(\eta_{i,j,t,1}, \eta_{i,j,t,2},\eta_{i,j,t,3} \right) = \log \big[ 1 + \exp \left\{ \eta_{i,j,t,1} \right\} + \exp \left\{ \eta_{i,j,t,2} \right\}\\
+ \exp \left\{ \eta_{i,j,t,1} + \eta_{i,j,t,2} + \eta_{i,j,t,3}\right\} \big],
\end{multline*}
and the parameters $\eta_{i,j,t,1} = f_{i,j,1}(\bfY_{t-1})$, $\eta_{i,j,t,2} = f_{i,j,2}(\bfY_{t-1})$ and $\eta_{i,j,t,3} = f_{i,j,3}(\bfY_{t-1})$ depend on time only through $\bfY_{t-1}$.  Note that $\eta_{i,j,t,3}$ controls the level of dependence between $y_{i,j,t}$ and $y_{j,i,t}$.  For example, $\eta_{i,j,t,3} = 0$ implies that $ y_{i,j,t}$ and $y_{j,i,t}$ are conditionally independent with $\Pr(y_{i,j,t} = 1 \mid \bfY_{t-1}) = \exp\left\{ \eta_{i,j,t,1} \right\}/\left(1 + \exp\left\{ \eta_{i,j,t,1} \right\} \right)$ and $\Pr(y_{j,i,t} = 1 \mid \bfY_{t-1}) = \exp\left\{ \eta_{i,j,t,2} \right\}/\left(1 + \exp\left\{ \eta_{i,j,t,2} \right\} \right)$. On the other hand, $\eta_{i,j,t,3} > 0$ favors outcomes in which $y_{i,j,t} = y_{j,i,t}$ (a phenomenon often called positive reciprocity), while $\eta_{i,j,t,3} < 0$ favors situations in which $y_{i,j,t} \ne y_{j,i,t}$ (often called negative reciprocity).  Hence, by allowing the values of $y_{i,j,t}$ and $y_{j,i,t}$ to be potentially correlated the model can accommodate (intra-temporal) reciprocity in the network.  

A full specification of the model requires that we specify the form of the functions $f_{i,j,1}$, $f_{i,j,2}$ and $f_{i,j,3}$.  A tempting option is to make these predictors dependent of all entries of $\bfY_{t-1}$, including all high order interactions.  However, such an approach leads to models with an extremely high number of parameters that is computationally unmanageable even for networks with a relatively small number of nodes.  On the other hand, while focusing only on first order effects associated with the entries of $\bfY_{t-1}$ can substantially reduce the number of parameters, the resulting model ignores interactions that could be expected to be important.  We take a middle ground approach and include in the specification of the functions $f_{i,j,1}$, $f_{i,j,2}$ and $f_{i,j,3}$ a subset of the first and second order effects that are associated with the interactions of nodes $i$ and $j$ among themselves and with other nodes during the previous period.  In particular, we set
\begin{multline}\label{eq:g1}
f_{i,j,l} (\bfY_{t-1})=\alpha_{i,j,l} + \beta_{i,j,l} y_{i,j,t-1} + \gamma_{i,j,l} y_{j,i,t-1}  \\
+ \sum_{k \neq i,j} \delta_{i,j,k,l} y_{i,k,t-1} + \sum_{k\neq i,j}\phi_{i,j,k,l}y_{k,j,t-1}  \\
+ \sum_{k \neq i,j} \psi_{i,j,k,l} y_{j,k,t-1} + \sum_{k\neq i,j}\omega_{i,j,k,l}y_{k,i,t-1}  \\
+ \rho_{i,j,l} y_{i,j,t-1}y_{j,i,t-1} + \sum_{k \neq i,j} \xi_{i,j,k,l} y_{i,k,t-1}y_{k,j,t-1}  \\
+ \sum_{k \neq i,j} \zeta_{i,j,k,l} y_{j,k,t-1}y_{k,i,t-1}
\end{multline}
for $l=1,2,3$.  To better motivate this specification, consider for example the structure of $f_{i,j,1} (\bfY_{t-1})$ in \eqref{eq:g1}.  As we showed before, we can roughly interpret $f_{i,j,1} (\bfY_{t-1})$ as controlling the probability of a directed link between $i$ and $j$.  Hence, $\alpha_{i,j,1}$ can be interpreted as the baseline probability of a link between nodes $i$ and $j$, the coefficients $\beta_{i,j,1}$ and $\gamma_{i,j1}$ can be interpreted as the persistence (momentum) in the relationship (e.g., if $\beta_{i,j,1} > 0$ then once trader $i$ starts selling to trader $j$, they tend to keep selling in future periods), the coefficients $\{ \delta_{i,j,k,1} : k \ne i,j \}$, $\{ \phi_{i,j,k,1} : k \ne i,j \}$, $\{ \psi_{i,j,k,1} : k \ne i,j \}$ and $\{ \omega_{i,j,k,1} : k \ne i,j \}$ capture diversification effects (e.g., if $\delta_{i,j,k,1} > 0$ then it is more likely that $i$ will sell to $j$ if it sold to $k$ in the previous term), $\rho_{i,j,1}$ captures inter-temporal reciprocity (as opposed to the intra-temporal reciprocity captured by $\eta_{i,j,t,3}$), and $\{ \xi_{i,j,k,1} : k \ne i,j \}$ and $\{ \zeta_{i,j,k,1} : k \ne i,j \}$ capture disintermediation effects (e.g., if $\xi_{i,j,k,1} > 0$ then so that $i$ is more likely to sell to $j$ if in the previous period $i$ sold to $k$ and $k$ sold to $j$, so that $i$ and $j$ tend to cut $k$ as middleman).

Although our model is not in the class of time-varying Exponential Random Graph models (tERGMs) \citep{HanFuXi10,CranDes11,SnijSteBunt10}, some classes of tERGMs can be obtained as special cases of our model.  Indeed, consider making the model parameters independent of the traders' identities so that $\alpha_{i,j,l} = \alpha_{l}$, $\beta_{i,j,l}=\beta_{l}$, $\gamma_{i,j,l} = \gamma_{l}$, $\delta_{i,j,k,l} = \delta_{l}$, $\phi_{i,j,k,l} = \phi_{l}$, etc.\ for all $i,j,k$.  In that case, the joint distribution $p( \bfY_2, \ldots, \bfY_T \mid \bfY_1 )$ is proportional to 
\begin{multline}\label{eq:tergm_equiv}
\exp\Bigg \{ \sum_{t=2}^{T} \sum_{l=1}^{3} \big[ \alpha_l S_{\alpha, l}(\bfY_t) + \beta_{l} S_{\beta, l}(\bfY_t, \bfY_{t-1})   \\
+ \gamma_{l} S_{\gamma, l}(\bfY_t, \bfY_{t-1}) + \delta_{l} S_{\delta, l}(\bfY_t, \bfY_{t-1}) \\
+ \phi_{l} S_{\phi, l}(\bfY_t, \bfY_{t-1}) + \psi_{l} S_{\beta, l}(\bfY_t, \bfY_{t-1})  \\
+ \omega_{l} S_{\omega, l}(\bfY_t, \bfY_{t-1})  + \rho_{l} S_{\rho, l}(\bfY_t, \bfY_{t-1}) \\
+ \xi_{l} S_{\xi, l}(\bfY_t, \bfY_{t-1})  + \zeta_{l} S_{\zeta, l}(\bfY_t, \bfY_{t-1}) \big] \Bigg\}
\end{multline}
where $S_{\alpha, l}(\bfY_t, \bfY_{t-1})$, $S_{\beta, l}(\bfY_t, \bfY_{t-1})$, etc.\ are appropriately chosen sufficient statistics, e.g.,
\begin{align*}
S_{\alpha, 1}(\bfY_t) &= \sum_{i=1}^{n}\sum_{j=i+1}^{n}  y_{i,j,t}   , \\
S_{\alpha, 2}(\bfY_t) &= \sum_{i=1}^{n}\sum_{j=1}^{i-1} y_{i,j,t}   ,  \\
S_{\alpha, 3}(\bfY_t) &= \sum_{i=1}^{n}\sum_{j=1}^{i-1} y_{i,j,t}y_{j,i,t}   ,  \\
S_{\delta, 2}(\bfY_t, \bfY_{t-1}) &= \sum_{i=1}^{n}\sum_{j=1}^{i-1} y_{i,k,t-1} y_{i,j,t} \\
S_{\xi,1}(\bfY_t, \bfY_{t-1}) & = \sum_{i=1}^{n}\sum_{j=i+1}^{n}  \sum_{k \ne i,j} y_{i,j,t} y_{i,k,t-1} y_{k,j,t-1} .
\end{align*}
Hence, \textit{if we were to strip away the identity of the nodes in the definition of the model coefficients}, the model would reduce to a tERGM constructed on the basis of sufficient statistics that correspond to the number of links in the network as well as the number of (some selected types of) two-stars and triangles.  By allowing the parameters to differ according to the identity of the nodes, our formulation generalizes the basic tERGM and allows for additional expressive power.


It is worthwhile noting that the collapsed model in \eqref{eq:tergm_equiv} does not include triangles in which all observations happen in the same time point. Hence, our model cannot capture the effects of \textit{intra-temporal} transitivity (i.e., an increase/decrease in the probability of a link between nodes $i$ and $j$ at time $t$ if they both link to a third node $k$ also at time $t$) on the evolution of the network. This modeling choice is made out of practical necessity; including this type of interactions into the model would complicate computation.  Indeed, assuming conditional independence among pairs of dyads on the same point in time is key to obtain a closed-form structure for the normalizing constant of the likelihood, which is in turn key to speed-up computation (see Section \ref{sec:computation}).  However, potential concerns surrounding this choice are mitigated by the fact that the second order interactions in \eqref{eq:g1} do allow us to capture \textit{inter-temporal} transitivity (i.e., an increase/decrease in the probability of a link between nodes $i$ and $j$ at time $t$ if they both linked to a third node $k$ at time $t-1$), which is more interesting and realistic in this type of scenarios.

\subsection{A penalized regression model}

The total number of parameters in our model is $\frac{n (n-1)}{2} \left\{ 9 + 18(n-2) \right\}$, which will typically be quite large.  In fact,  the number of parameters in the model will often be larger than the number of observations available to estimate them.  To address this issue we adopt a regularized likelihood approach based on $L^1$ penalty functions.  More specifically, point estimates for the model parameters are obtained by solving:
\begin{align}\label{eq:optim}
\arg\max_{\bfalpha, \bfTheta} \sum_{i < j} \left\{  V_{i,j}(\bfalpha_{i,j}, \bfTheta_{i,j}) - \lambda \|\bfTheta_{i,j}\|_{1} \right\}
\end{align}
where 
\begin{multline*}
V_{i,j}(\bfalpha_{i,j}, \bfTheta_{i,j}) = 
\sum_{t=2}^{T}
\big\{  
y_{i,j}\left[ \alpha_{i,j,1} + \bfx_{i,j,t}^{T} \bftheta_{i,j,1} \right] +
y_{j,i}\left[ \alpha_{i,j,2} + \bfx_{i,j,t}^{T} \bftheta_{i,j,2} \right] + \\
y_{i,j}y_{j,i}\left[ \alpha_{i,j,3} + \bfx_{i,j,t}^{T} \bftheta_{i,j,3} \right]
\big\}
\end{multline*}
is the (unpenalized) log-likelihood, $\| \cdot \|_{1}$ denotes the $L_{1}$-norm, $\lambda> 0$ is the penalty parameter controling the shrinkage level of the coefficients towards zero, $\bfalpha_{i,j}= (\alpha_{i,j,1} , \alpha_{i,j,2} , \alpha_{i,j,3})'$, the vector of parameters $\bftheta_{i,j,r}$ is defined as
\begin{multline*}
\bftheta_{i,j,r} = (\beta_{i,j,r}, \gamma_{i,j,r}, \delta_{i,j,1,r}, \ldots, \delta_{i,j,n,r},
\phi_{i,j,1,r}, \ldots, \phi_{i,j,n,r},  \psi_{i,j,1,r}, \ldots, \psi_{i,j,n,r},  \\ 
\omega_{i,j,1,r}, \ldots, \omega_{i,j,n,r}, \rho_{i,j,r}, \xi_{i,j,1,r}, \ldots, \xi_{i,j,n,r},
\zeta_{i,j,1,r}, \ldots, \zeta_{i,j,n,r} )'   ,
\end{multline*}
$\bfTheta_{i,j}=\left( \bftheta_{i,j,1}',\bftheta_{i,j,2}',\bftheta_{i,j,3}' \right)'$, and the vector of covariates is
\begin{multline*}
\bfx_{i,j,t} = (y_{i,j,t-1}, y_{j,i,t-1}, y_{i,1,t-1}, \ldots, y_{i,n,t-1},\\
y_{1,j,t-1}, \ldots, y_{n,j,t-1}, y_{j,1,t-1}, \ldots, y_{j,n,t-1}, y_{1,i,t-1}, \ldots, y_{n,i,t-1}, \\
y_{i,j,t-1}y_{j,i,t-1},y_{i,1,t-1}y_{1,j,t-1}, \ldots, y_{i,n,t-1}y_{n,j,t-1},\\
y_{j,1,t-1}y_{1,i,t-1}, \ldots, y_{j,n,t-1}y_{n,i,t-1})'  .
\end{multline*}
Note that the structure $V_{i,j}$ is equivalent to that of a multinomial likelihood and that the intercept parameters $\left\{\alpha_{i,j,r}\right\}$ remain unpenalized.  Furthermore, the imposition of a lasso penalty is equivalent to assuming independent double exponential prior distributions with variance $2/\lambda$ on each component of $\bfTheta$, so that the point estimates obtained from \eqref{eq:optim} coincide with the maximum a posteriori (MAP) estimates.

\section{Estimation and prediction}\label{sec:computation}

One important consequence of the conditional independence assumption is that \eqref{eq:optim} can be broken down into $n(n-1)/2$ optimization problems, each one corresponding to fitting a separate $L^1$ regularized multinomial regression for each pair of nodes in the network.   In the sequel we focus on an algorithm to solve each of these independent problems (and drop the subindex $(i,j)$ to simplify notation). This algorithm is then implemented in a parallel environment.

There is an extensive literature on efficient algorithm for estimation in $L^1$ regularized multinomial regression (e.g., see \citealp{Krishna05}, \citealp{GoldOsher09}, and \citealp{FriHaTib10}).  In this paper we resort to a relatively simple computational algorithm similar to iterative reweighed least squares (see \citealp{FriHaTib10}).  In particular, we solve \eqref{eq:optim} by iteratively setting 
\begin{align*}
\left( \hat{\bfalpha}^{(m+1)}, \hat{\bfTheta}^{(m+1)} \right) = \underset{\bfTheta} \argmax \, Q \left(\bfalpha, \bfTheta \mid \hat{\bfalpha}^{(m)},\hat{\bfTheta}^{(m)}\right)
\end{align*}
until convergence, where $Q(\bfalpha, \bfTheta \mid \tilde{\bfalpha}, \tilde{\bfTheta})$ is a surrogate function obtained by replacing $V(\bfalpha, \bfTheta)$ by its second-order Taylor expansion around the current iterate.  Furthermore, rather than attempting to solve the problem using blockwise updates, we proceed with componentwise steps. In particular, the estimate of a component $\theta_{r,k}$ of $\bftheta_{r}$ is updated as
\begin{align}
\hat{\theta}^{(m+1)}_{r,k}=\text{soft}\left(\hat{\theta}^{(m)}_{r,k}-\frac{g^{(m)}_{r,k}}{G^{(m)}_{r,k,k}};\frac{-\lambda}{G^{(m)}_{r,k,k}} \right),
\end{align}
where $\text{soft}(w,\lambda)=\sign(w)\max\{0,|w|-\lambda\}$ is the soft-thresholding operator, $g^{(m)}_{r,k} = \frac{\partial V}{\partial \theta_{r,k}}$ and $G^{(m)}_{r,k,k} = - \frac{\partial^2 V}{\partial \theta^{2}_{r,k}}$ are the gradient and the information in the direction of $\theta_{r,k}$ evaluated in the current iterate values $\hat{\bfalpha}^{(m)} and \hat{\bfTheta}^{(m)}$. On the other hand, since the intercepts are not penalized, their estimates are updated using the recursion
$$
\hat{\alpha}^{(m+1)}_{k} =\hat{\alpha}^{(m)}_{k}-\frac{g^{(m)}_{k}}{G^{(m)}_{k,k}} .
$$

One consequence of the use of $L_1$ penalized likelihoods is that point estimates of the coefficients can be exactly zero. Hence, the algorithm automatically performs variable selection, allowing us to assess the presence of diversification and disintermediation effects in the network. This allows us to explicitly test hypotheses about the kind of effects that influence the evolution of the network.  However, when the regression matrix is not full rank (for example, when  $T < 9 + 18(n-2)$), interpretation of the individual effects is difficult because of confounding/multicolinearity. To address this issue we focus on identifying regression coefficients for which there is no evidence of significance. These are selected by identifying the effects that lie in the orthogonal complement of the column space of $\bfX_{i,j}(\mathcal{A_{\lambda}})$, the submatrix that contains the columns associated with variables that have been identified as significant using the penalty $\lambda$.

\subsection{Selection of the penalty parameter}\label{se:select}

The value of the penalty $ \lambda$ has a direct impact on the quality of the estimates and predictions generated by the model.  Our default approach is to select $\lambda$ from among a pre-specified grid of values by maximizing the Bayesian Information Criteria (BIC)
\begin{align*}
BIC_{\lambda} =\sum_{i < j} \left[ 2V_{i,j}(\hat{\bfalpha}_{i,j}, \hat{\bfTheta}_{i,j}) - \mathcal{K}_{i,j}(\lambda) \log (T-1)\right]  ,
\end{align*}
where $\mathcal{K}_{i,j}(\lambda) = \text{rank}\left\{ \bfX_{i,j}(\mathcal{A}_{\lambda}) \right\}$ is an estimate of the number of degrees of freedom when the penalty parameter $\lambda$ is used to compute $(\hat{\bfalpha}_{i,j}, \hat{\bfTheta}_{i,j})$, and $\bfX_{i,j}(\mathcal{A}_{\lambda})$ is a $(T-1)\times d$ matrix whose $t$-th row contains a subset of elements of $\bfx_{i,j,t}$ and whose columns correspond to the covariates for which $\hat{\bftheta}_{i,j,r}$ is different from zero for at least one value $r=1,2,3$ \citep{ParkHas06,ZouHasTib07,TibTay12}.  Note that for all values of $\lambda$, the degrees of freedom satisfy the condition $ 0 \leq \mathcal{K}_{\lambda}\leq \text{min}\{d,T-1\}$.

\subsection{Link Prediction}

Although the main goal of our analysis is not short-term link prediction, our autologistic model can be used for this purpose. In particular, given a point estimate $(\hat{\bfalpha}_{i,j}, \hat{\bfTheta}_{i,j})$ based on an observed sample $\bfY_1, \ldots, \bfY_T$, we can estimate (for $i>j$) the probability of a directed link from node $i$ to node $j$ at time $T+1$ as
\begin{multline*}
\hat{p}\left(y_{i,j,T+1} = 1 \mid \bfY_{T} \right) = p\left[(y_{i,j,T+1},y_{j,i,T+1}) = (1,0) \mid \hat{\bfalpha}_{i,j}, \hat{\bfTheta}_{i,j})\right]\\
+ p\left[(y_{i,j,T+1},y_{j,i,T+1}) = (1,1) \mid \hat{\bfalpha}_{i,j}, \hat{\bfTheta}_{i,j})\right],
\end{multline*}  
with a similar expression being valid for $\hat{p}\left(y_{j,i,T+1} = 1 \mid \bfY_{T} \right)$. 

\subsection{Theoretical properties}

Because our estimation procedure reduces to fitting independent $L^1$ regularized multinomial logistic regressions for each pair of nodes, the procedure shares all the positive (and negative) properties of this type of approaches (for a review, see \citealp{VidaBi13}). For example, for $n$ fixed and $T$ growing to infinity, the estimators are both consistent and sparse consistent, and have the oracle property (see \citealp{fan2001variable} for the full rank case and \citet{ZhaoYu06} and \citealp{LeeSunTaylor14} for the non-full rank case).  Under some additional restrictions (such as the Restricted Eigenvalue (RE) condition described in \citealp{Bickel09}), these results apply for $n$ growing with $T$ (e.g., see \citealp{Wainwright09} and \citealp{KakadeSST10}).

\section{Application}\label{sec:applications}

\subsection{Simulation study} \label{sec:simulation}

We begin by demonstrating the predictive performance of the model in a simulated dataset consisting of $T=201$ networks observed over $n=71$ nodes (the same values as in the NYMEX data). The data was generated according to our model in such a way that for each of the 2,485 pairs of nodes only six non-zero coefficients are present for each class, $l=1,2,3$. Three of the non-zero coefficients for each class correspond to ${\alpha}_{i,j,l}$, ${\beta}_{i,j,l}$, and ${\gamma}_{i,j,l}$. We randomly draw these parameters from common Gaussian distributions across pairs (e.g. ${\alpha}_{i,j,l} \sim N(\bar{\alpha}_{l},  \tau^{2}_{l})$). The other relevant coefficients correspond to ${\xi}_{i,j,k,l}$ for three different values of $k$. Four groups of pairs of traders of similar sizes were simulated with different selections of the three values of $k$, and the respective parameter values of ${\xi}_{i,j,k,l}$ where fixed equal within each of the groups and with opposites signs to the global mean of ${\beta}_{i,j,l} $ and ${\gamma}_{i,j,l} $. Under this simulation scheme, the persistence of the relationship between the nodes $i$ and $j$ and a few transitive relationships drive the network structure and dynamics over time. The resulting network is relatively dense with a average number of links of 2971 (out of 4970 possible ties), and it shows low reciprocity and high transitivity. 

\begin{figure}
\begin{center}
\includegraphics[scale=0.35,keepaspectratio]{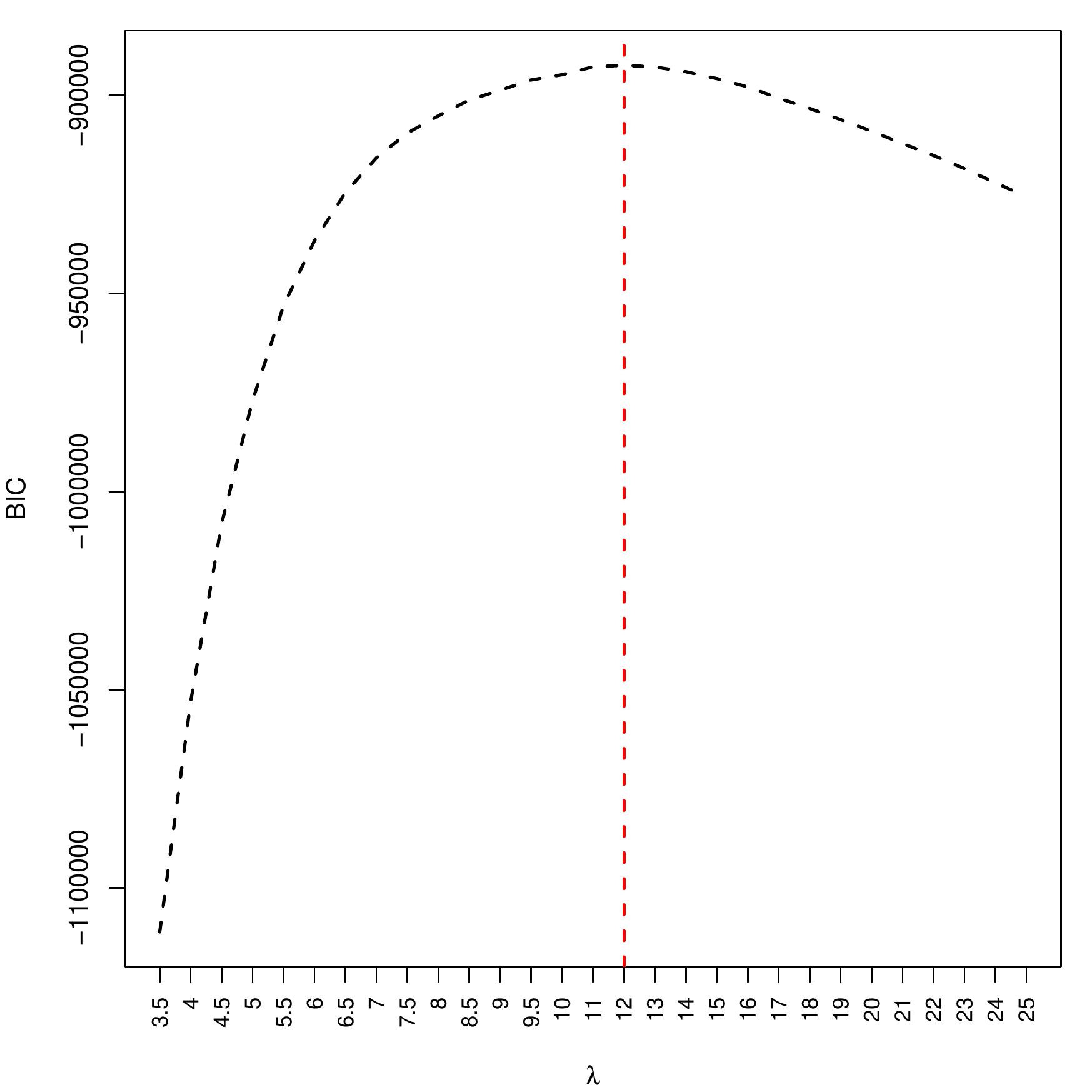}
\end{center}
\caption{BIC values over a grid of values of $\lambda$ for autologistic model in simulated dataset.}\label{fig:lambdaSim}
\end{figure}

As a benchmark, we also fit to the data a tERGM \citep{LeCranDes14} that includes all the typical ERGM terms, the square root of in and out-degrees as node covariates, and the lagged network and the delayed reciprocity to model cross-temporal dependencies. This model was fit using the \texttt{btergm} function of the \texttt{R} package \texttt{xergm}. Our evaluation is based on an out-of-sample crossvalidation exercise where we held out the last ten weeks in the data set and made one-step-ahead predictions for the structure of the held-out networks. More specifically, for each $t=191,192,\ldots,200$ we use the information contained in $\bfY_1, \ldots, \bfY_{t}$ to estimate the model parameters and obtain predictions for $\hat{\bfY}_{t+1}$. The quality of the prediction is evaluated by constructing receiver operating characteristic (ROC) curves and computing the area under this curve. Predictions for the tERGM model are based on 1,000 MCMC simulations generated using the default parameter values for the \texttt{xergm} package (see \textsf{btergm} documentation for more details).

\begin{figure}
\begin{center}
\includegraphics[scale=0.35,keepaspectratio]{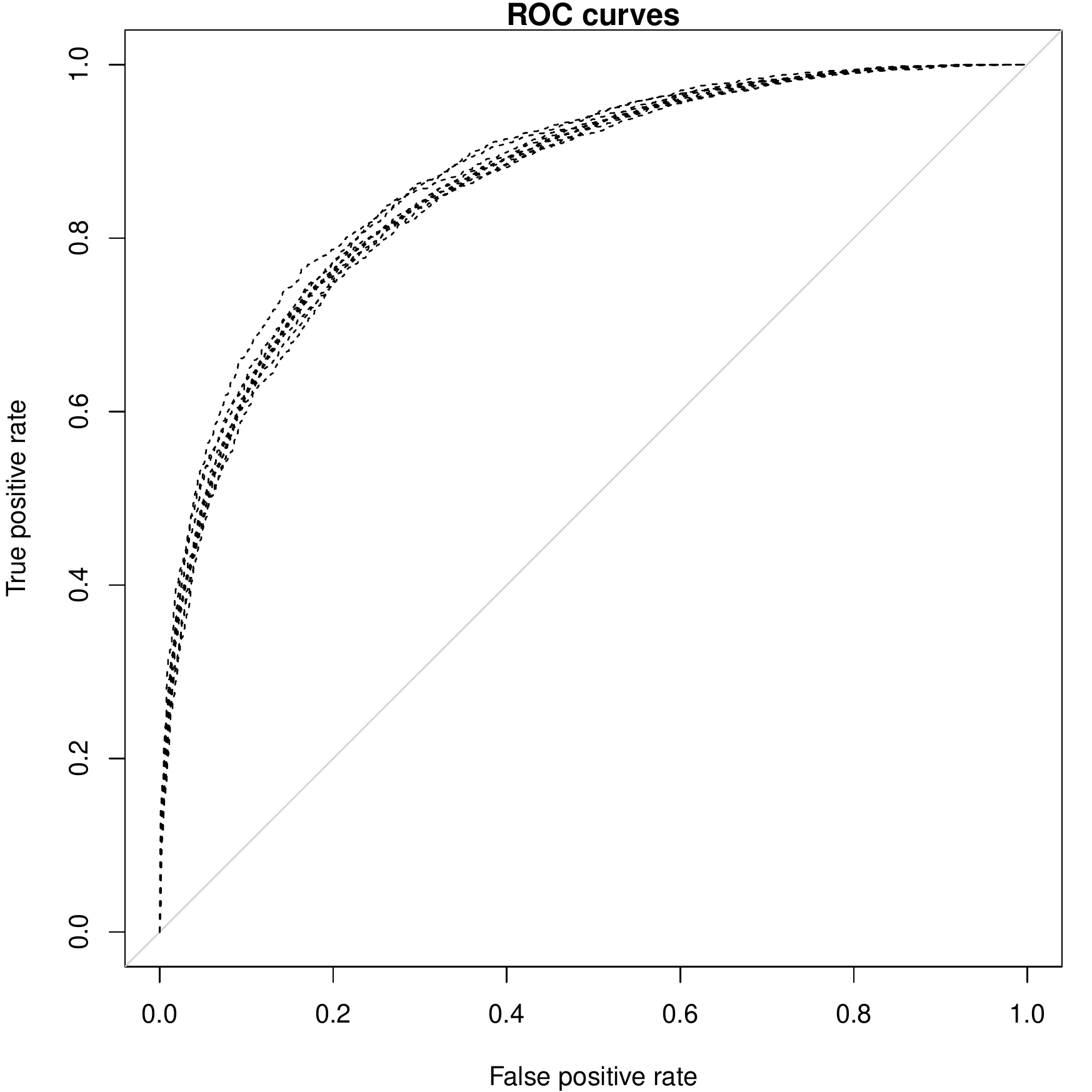}
\includegraphics[scale=0.35,keepaspectratio]{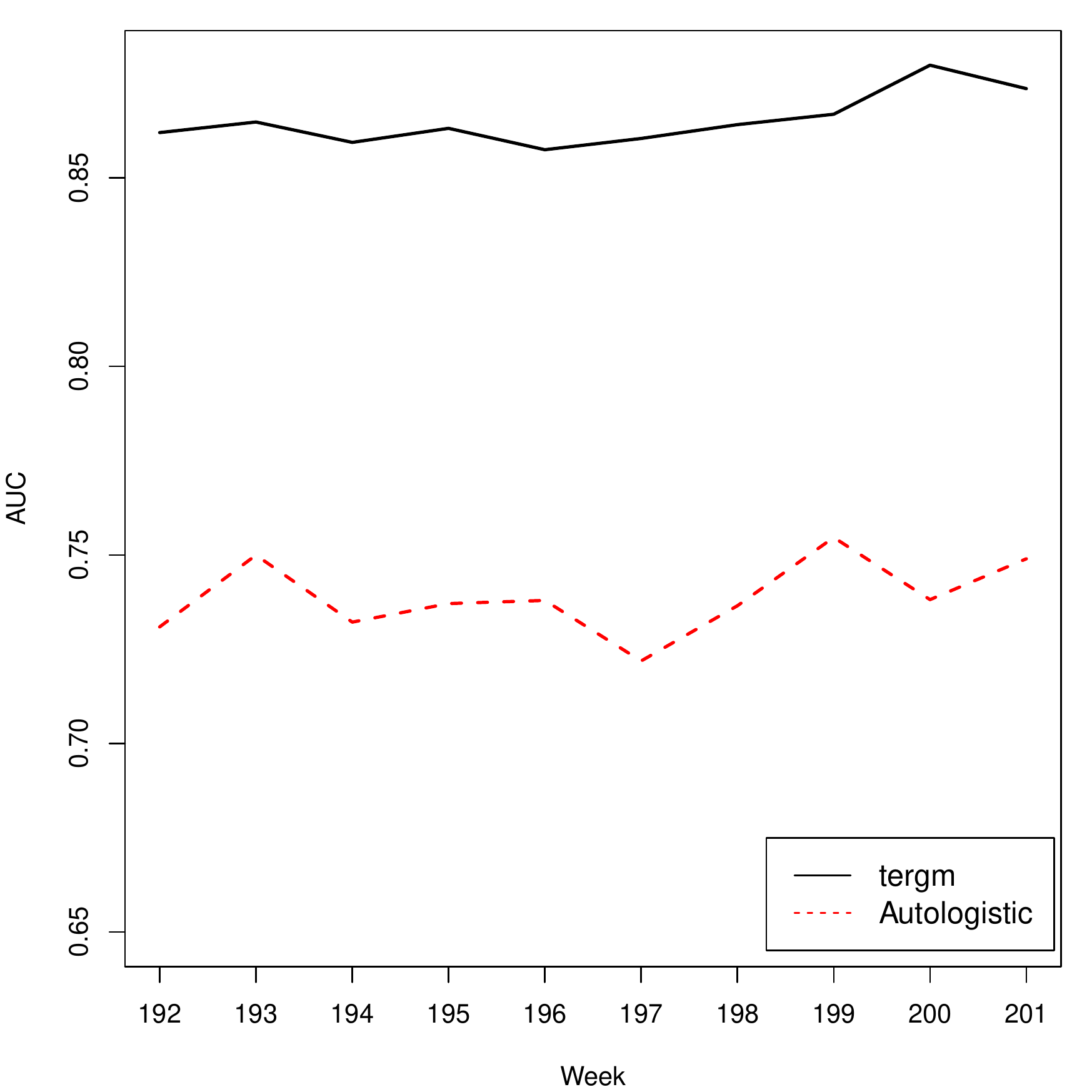}
\end{center}
\caption{Plots of the ten operating characteristic curves associated with one-step-ahead out of sample predictions from the autologistic model, and the area under the curves (AUC) for the autologistic model and the temporal ERGM for the simulated dataset.}\label{fig:simula}
\end{figure}

We search for the optimal value of $\lambda$ over a grid of 29 values between 3.5 and 25, with the optimal value being $\lambda=12$ (see Figure \ref{fig:lambdaSim}). Figure \ref{fig:simula} shows the ten operating characteristic curves associated with one-step-ahead out of sample predictions from our autologistic model, along with estimates of the area under the receiver operating characteristic curves (AUC) for the proposed model and the tERGM.  From these results it can be seen that the predictive accuracy of the temporal ERGM is poor as it only reaches AUC values below 75\% in most cases.  In this scenario, the autologistic model shows superior prediction ability outperforming the tERGM by 11\% to 14\% in the AUC values for all cases.
  
\subsection{Analysis of the NYMEX data}\label{sec:NYMEX}

In this section we analyze a sequence of $T=201$ weekly financial trading networks constructed from proprietary trades  in the natural gas futures market on the New York Mercantile Exchange (NYMEX) between January 2005 and December 2008 (see Section \ref{se:data}). Previous exploration of this data showed that these trading networks are moderately sparse (with an average of 826 links out 4,970 possible ones), and consistently show very high reciprocity, moderate transitivity, mixing patterns and community structure \citep{BetaRodBoyd15}. These features suggests that link formation between a pair of nodes is very likely to depend on how other actors relate in the network. 

\begin{figure}
\begin{center}
\includegraphics[scale=0.35,keepaspectratio]{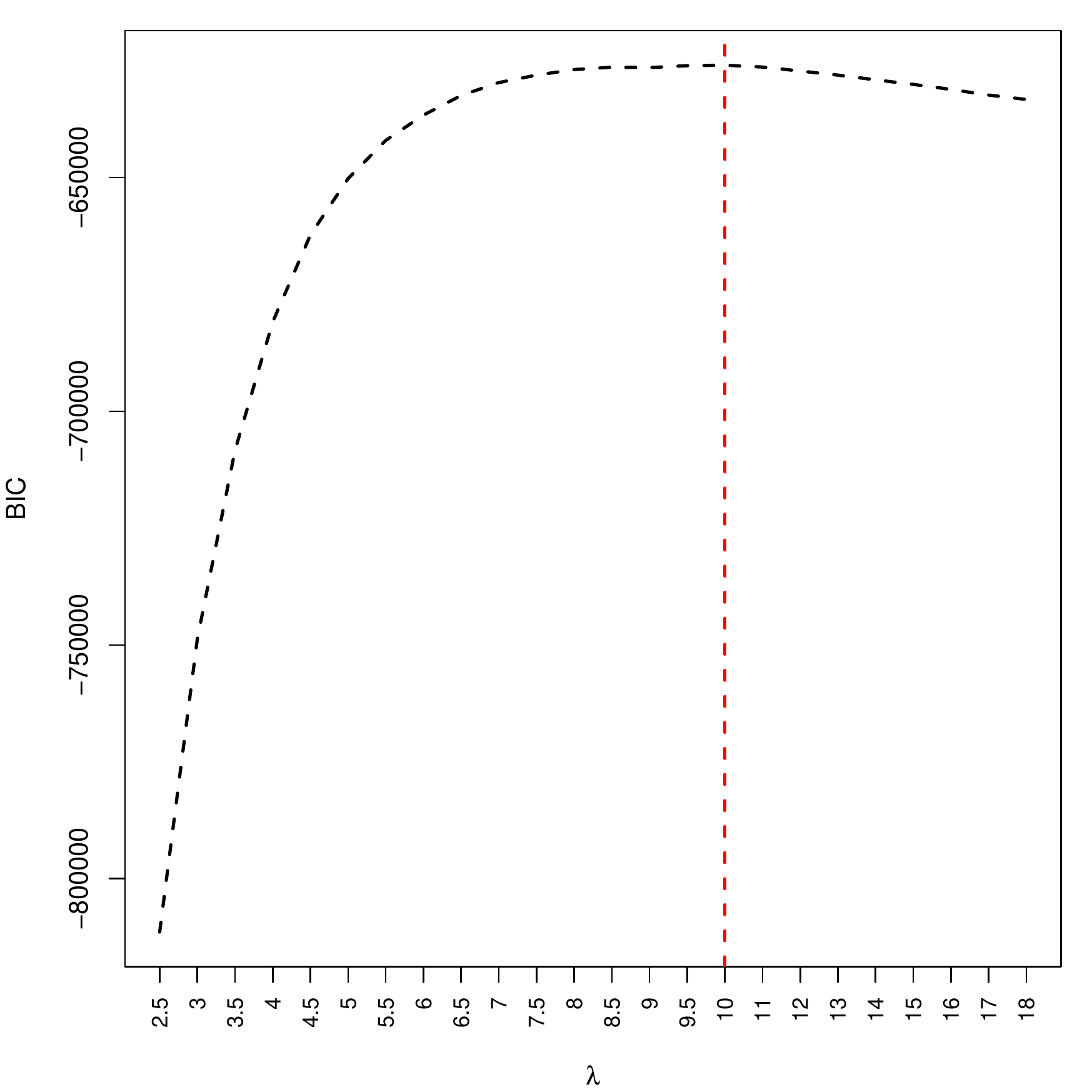}
\end{center}
\caption{BIC values over a grid of values of $\lambda$ for autologistic model in NYMEX financial trading network.}\label{fig:lambdaTrad}
\end{figure}
  
\begin{figure}
\begin{center}
\includegraphics[scale=0.35,keepaspectratio]{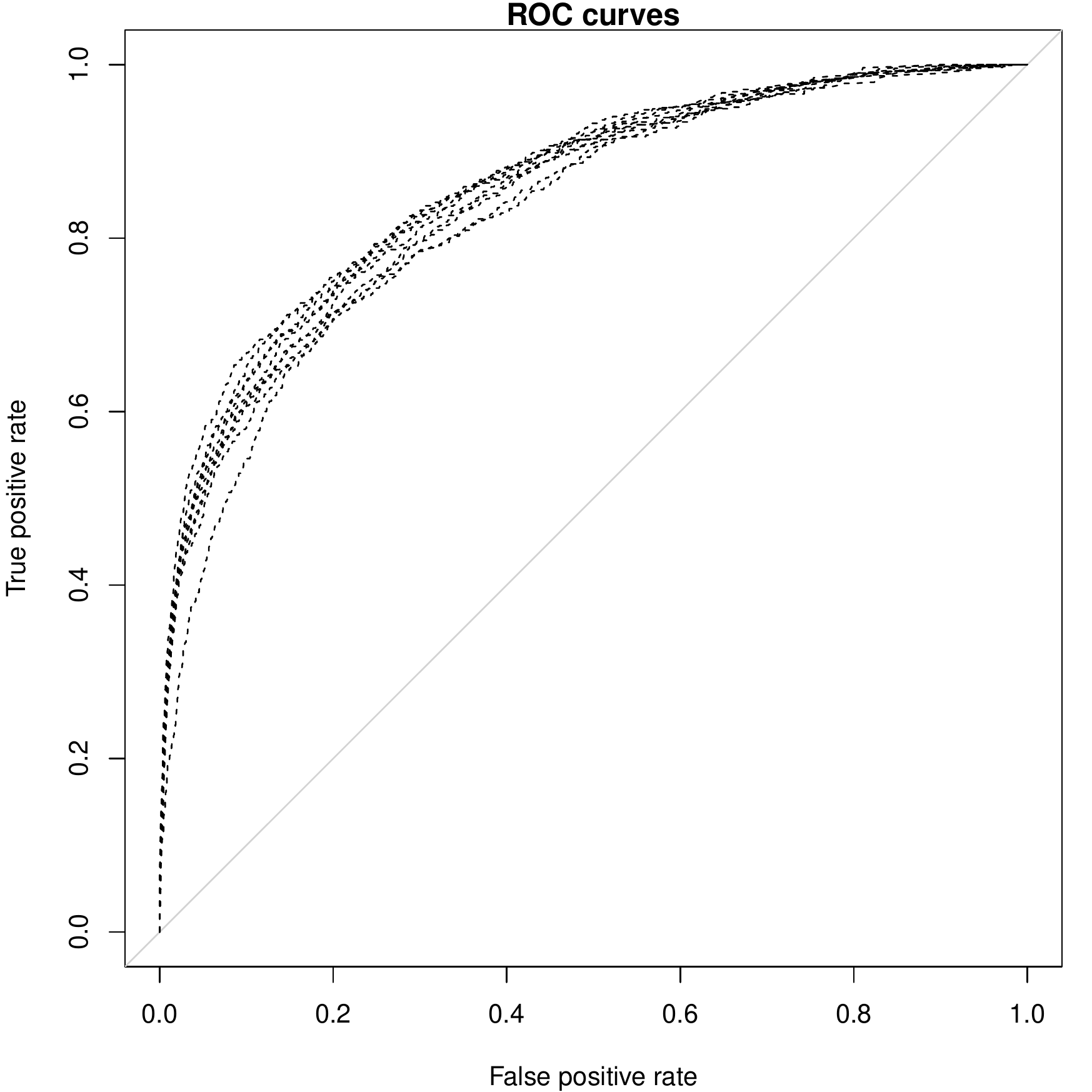}
\includegraphics[scale=0.35,keepaspectratio]{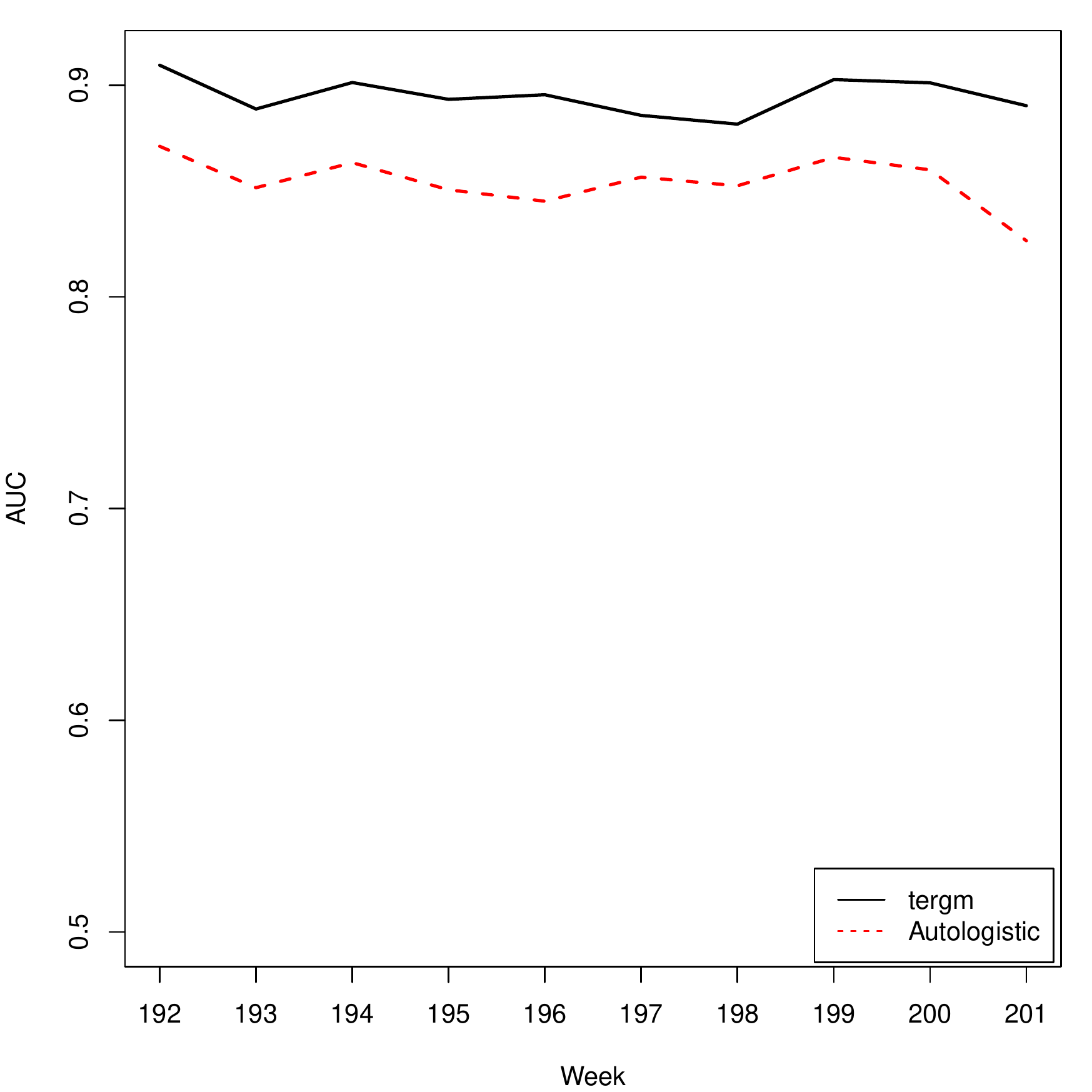}
\end{center}
\caption{Plots of the ten operating characteristic curves associated with one-step-ahead out of sample predictions from the autologistic model, and the area under the curves (AUC) for the autologistic model and the temporal ERGM for the NYMEX financial trading network.}\label{fig:traders}
\end{figure}

Selection of the optimal penalization parameter in this case was performed by searching over a grid of 24 values between 2.5 and 18 for a resulting optimal 
 value of $\lambda=10$ (see Figure \ref{fig:lambdaTrad}). As before, we carry out an out-of-sample cross-validation exercise in which our model and the tERGM are fitted to the first 191 weeks, and the estimated model parameters are then used to predict each of the last 10 weeks of data. Figure \ref{fig:traders} shows the ten operating characteristic curves associated with the out-of-sample predictions from the autologistic model, and the estimates of the area under the receiver operating characteristic curves (AUC) for both models. In this particular case, the temporal ERGM outperforms our proposed model by between 3\% and 6\% in the AUC. However, the prediction accuracy of the autologistic model is resonably good with an average AUC value of 85\% over the 10 weeks. The temporal ERGM slightly outperforming our model for the NYMEX data is probably a result of the presence of moderate \emph{intra-temporal} transitivity that we are unable to capture in our model. In contrast, the simulated network in Section \ref{sec:simulation} is dominated by \emph{inter-temporal} transitivity that the temporal ERGM is unable to capture.
 
Now, we turn our attention to the interpretation of the regression coefficients. Recall that the trading network has $n=71$ traders, so that the autologistic model includes 1251 covariates (excluding the unpenalized intercept).  Of these 1251 coefficients, 6 capture persistence effects, 3 capture inter-temporal reciprocity, 828 capture substitution effects, and 414 capture disintermediation effects. However, since the number of covariates is much larger than the number of observations, these effects are confounded with each other, complicating the interpretation of the model. To address this issue we focus on identifying effects for which there is no evidence of significance (see Section \ref{sec:computation}).

\begin{figure}
\begin{center}
\includegraphics[scale=0.35,keepaspectratio]{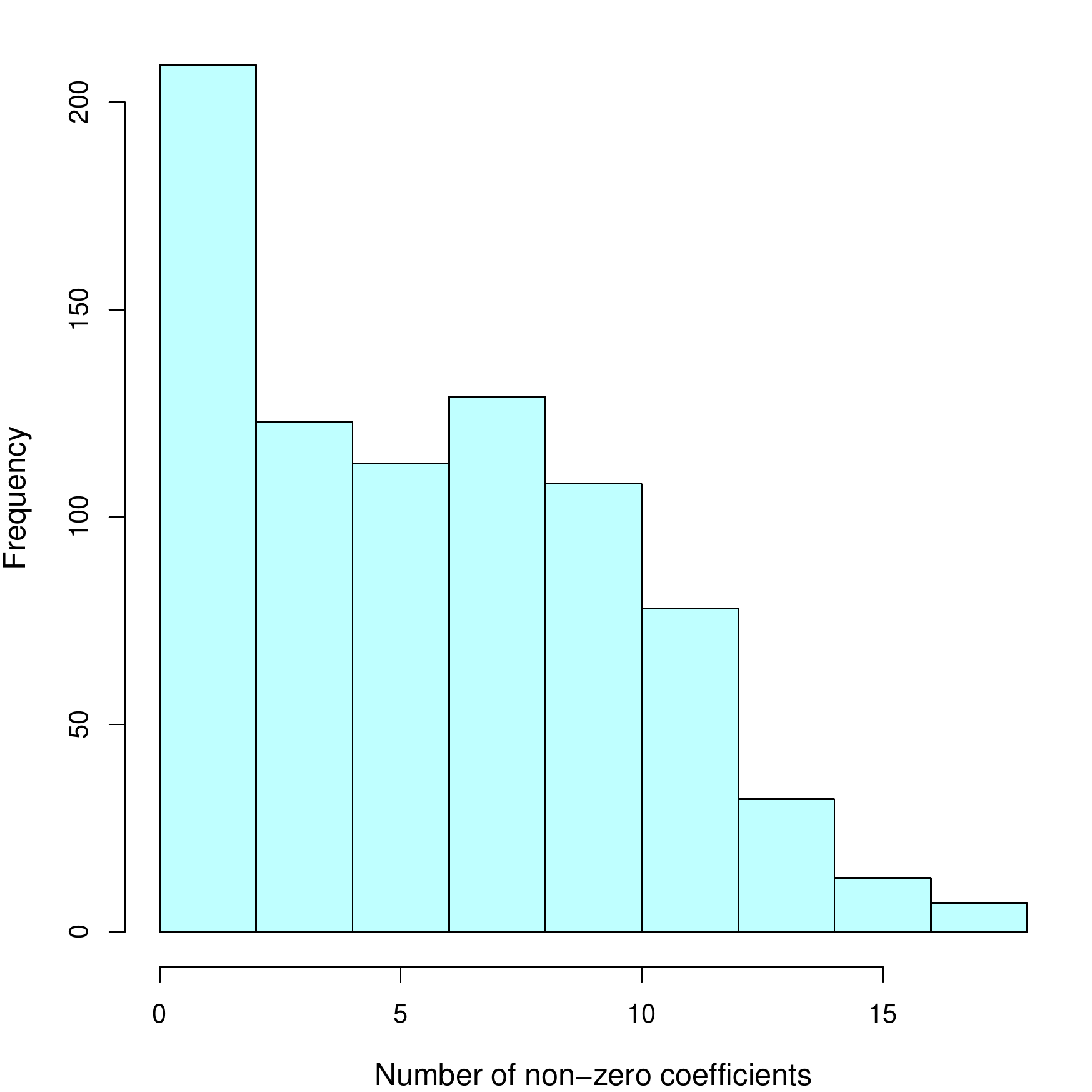}
\includegraphics[scale=0.35,keepaspectratio]{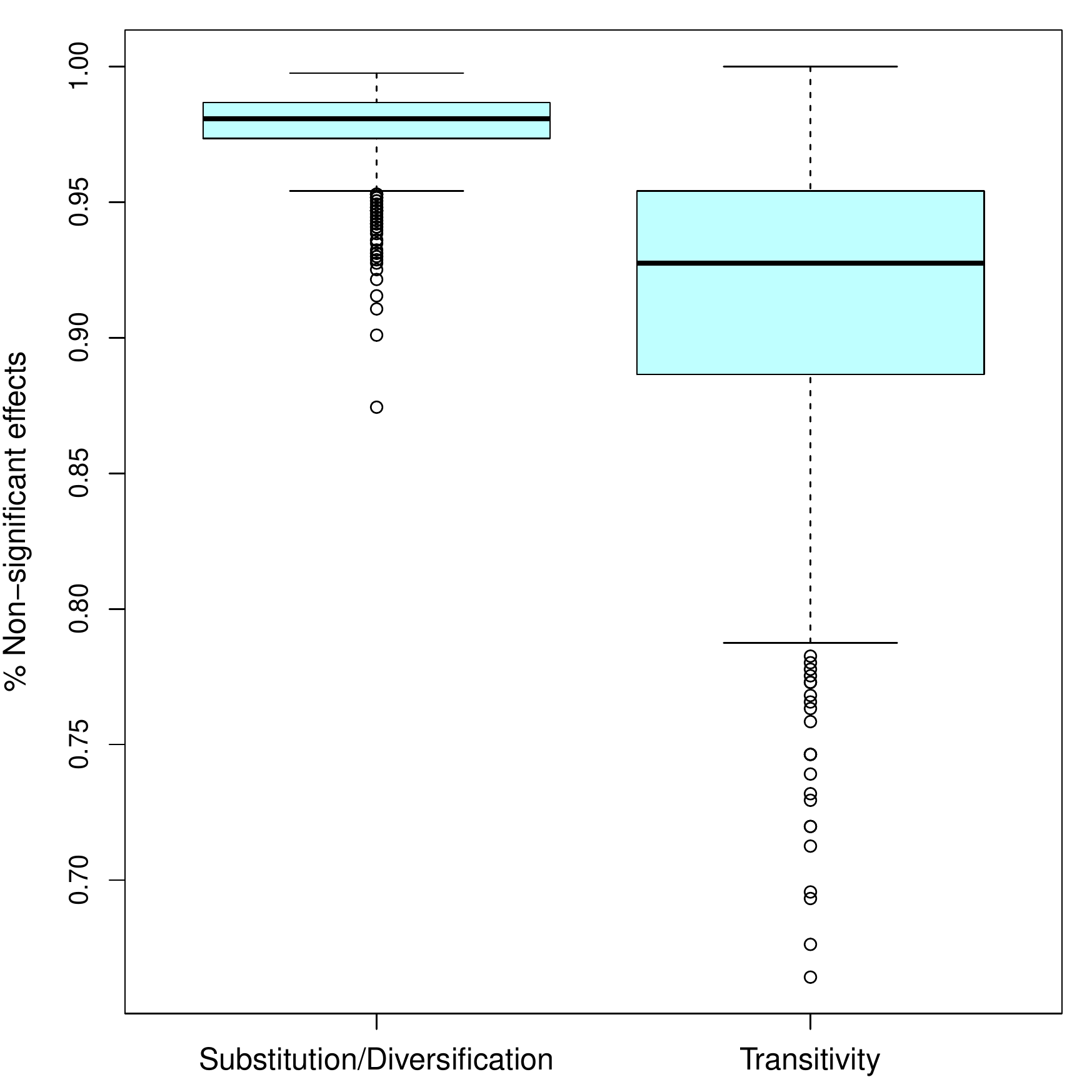}
\end{center}
\caption{Histogram of the number non-zero coefficients and boxplots of the percentage of non-significant effects of substitution/diversification and transitivity
 for 812 pairs of traders.}\label{fig:effects}
\end{figure}

First, we note that the individual regression models for each pair tend to be quite sparse. Indeed, only 812 out of 2485 regressions have at least one non-zero regression coefficient aside from the intercept and the number of non-zero coefficients (i.e., the rank of $\bfX_{i,j}(\mathcal{A_{\lambda}})$) tend 
to be very low across these pairs (see first panel Figure \ref{fig:effects}).  However, the number of significant effects varies dramatically across the different pairs. For example, the interaction between traders 2 and 17 (we identify traders by a number rather than their name because of confidentiality restrictions) seems to be driven by five significant effects: two of them are associated with persistence, one with reciprocity, and the other two with substitution/diversification. This is in contrast with the interaction between traders 64 and 71, which appear to be driven by over 100 potentially significant effects.
\begin{table}[h]
\caption{Number of effects, and non-presence percentage over 812 pairs of traders.} \label{tab:Effects}
\begin{center}
\begin{tabular}{lcc}
   &{\bf \# of effects}  &{\bf \% non-present} \\
\hline \hline
{\bf Persistence}   &  6 &  72.8 \\
{\bf Reciprocity}    & 3     & 88.8 \\                           
{\bf Substitution/Diversification}       & 828  & 0.0 \\
{\bf Disintermediation}   &414  & 0.4\\
\hline
\end{tabular}
\vspace{-0.5cm}
\end{center}
\end{table}
To understand the overall impact of different trading mechanisms we focus on the 812 pairs that show at least one significant effect and note that a large percentage have no persistence (72.8\%) or inter-temporal reciprocity (88.8\%) coefficients that are significant (see Table \ref{tab:Effects}).  In contrast, each one of these 812 pairs presents at least one substitution/diversification coefficient that might be significant, and the vast majority (99.6\%) present at least one potentially significant inter-temporal transitivity effect. This clearly suggests that second-order effects (substitution/diversification and transitivity) are much more important in this financial trading network than first order effects (persistence and reciprocity). 

To better understand the relative importance of the second order effects, we also compute the number of non-significant effects of each type for each pair of traders (see Figure \ref{fig:effects}). Note that, although all 812 pairs present at least one substitution/diversification effect, the number of these effects that might be significant on each pair is relatively small. In contrast, the number of potentially significant coefficients associated with inter-temporal transitivity effects tends to be larger, with a few pairs presenting more than 30\% of potentially significant effects. These results suggests that the evolution of this trading network is driven in majority by inter-temporal transitivity effects.

We can contrast these results with those obtained from the tERGM model (see table \ref{ta:tERGM}).  Interestingly, all coefficients appear to be significant in this case. Hence, unlike our model, the tERGM suggests that trading is ``sticky'' (both momentum effects are significant).  This is likely driven by the fact that in the default specification of the tERGM the momentum terms are the only ones that capture the temporal evolution of the network.
\begin{table}
\begin{center}
\begin{tabular}{lcc}
\textbf{Parameter}                      &   \multicolumn{2}{c}{\textbf{Estimate}}                     \\ \hline \hline
Number of edges                       &     -7.542   &   (-7.671,  -7.421) \\
Number of paths                        &      0.174   &   (0.122,  0.230) \\
Reciprocity                                 &      1.134   &   (1.103,  1.165) \\
Transitive Triplets                       &     -0.043   &   (-0.046,  -0.039) \\ 
Cyclic triplets                              &      0.074   &   (0.067,  0.081) \\
Sqrt of in-degree for receiver     &      1.096   &   (1.075,  1.117) \\
Sqrt of out-degree for receiver   &     -0.368   &   (-0.385,  -0.350) \\
Sqrt of out-degree for emitter     &      0.751   &   (0.727,  0.777) \\
Momentum                                 &      0.771   &   (0.745,  0.797) \\
Momentum of reciprocal            &       0.720   &   (0.696,  0.744) \\ \hline
\end{tabular}
\end{center}
\caption{Point estimates and 95\% confidence intervals for the coefficients of tERGM model in the NYMEX FTN data.}\label{ta:tERGM}
\end{table}

%
%
%

\section{Discussion}\label{sec:discussion}

We introduced a novel statistical model for the analysis of financial trading networks and applied it to study the NYMEX natural gas futures market between January 2005 and December 2008. Our analysis shows that diversification and substitution effects rather than persistence tend to dominate this market's microstructure.

Our approach focuses on $L_1$ penalties mainly because of computational expediency.  However, alternatives such as the adaptive Lasso \cite{zou2006adaptive} or smoothly clipped absolute deviation (SCAD) \citep{fan2001variable} penalties can potentially improve variable selection. We would also like to explore in the future fully Bayesian implementation using spike-and-slab priors (e.g., see \citealp{ishwaran2005spike}) instead of convex non-differentiable penalties. However, computation for this type of models (particularly for networks with a large number of nodes) is challenging.  Furthermore, we note that the model can be easily extended to undirected network by considering a reduced set of predictors, and to weighted networks by replacing the multinomial likelihood with an appropriate member of the exponential family.  Similarly, we could extend to model to consider higher order autoregressive processes, but the number of parameters grows dramatically in that case.  One way to deal with this issue is to focus on a smaller number of hyperparameters or drop interactions from the model.  

In addition to the componentwise maximization algorithm described in Section \ref{sec:computation}, we also investigated the use of a split-Bregman algorithm \citep{GoldOsher09} and a stochastic gradient algorithm \citep{ShalTew11}, but found both algorithms to have suboptimal performance in this problem.  However, for problems with very large $T$ (e.g., high frequency FTNs in highly liquid markets), a stochastic gradient descent approach  might be appropriate, and its implementation is relatively straightforward.

\section*{Acknowledgements}

This research was partially supported by NSF/DMS award number 1441433.

\bibliographystyle{plainnat}    
\bibliography{autolog}

\end{document}